\documentclass[prc,twocolumn,epsfig]{revtex4}
\usepackage{graphics}
\usepackage{epsfig}
\usepackage{amsfonts}
\usepackage{amsmath}
\usepackage{bm}

\usepackage{color}
\begin{document}
\title{Pairing and deformation effects in nuclear excitation spectra}
\author{A. Repko$^{1}$, J. Kvasil$^{2}$, V.O. Nesterenko$^{3,4}$,
and P.-G. Reinhard$^{5}$}
\affiliation{$^1$
Institute of Physics, Slovak Academy of Sciences, 84511 Bratislava, Slovakia}
\affiliation{$^2$
Institute of Particle and Nuclear Physics, Charles University,
CZ-18000 Prague 8, Czech Republic}
\affiliation{$^3$
Laboratory of Theoretical Physics,
Joint Institute for Nuclear Research, Dubna, Moscow Region, 141980, Russia}
\affiliation{$^4$
State University "Dubna", Dubna, Moscow Region, 141980, Russia}
\affiliation{$^5$
Institut f\"ur Theoretische Physik II, Universit\"at Erlangen, D-91058, Erlangen, Germany}
\date{\today}
\begin{abstract}
We investigate effects of pairing and of quadrupole
    deformation on two sorts of nuclear excitations,
  $\gamma$-vibrational $K^{\pi}=2^+$ states and dipole resonances
  (isovector dipole, pygmy, compression, toroidal). The analysis is
  performed within the quasiparticle random-phase approximation (QRPA)
  based on the Skyrme energy functional using the Skyrme
  parametrization SLy6. Particular attention is paid to i) the
  role of the particle-particle (pp) channel in the residual
    interaction of QRPA, ii) comparison of volume pairing (VP) and
  surface pairing (SP), iii) peculiarities of deformation splitting
  in the various resonances.  We find that the impact of
  the pp-channel on the considered excitations is negligible.
  This conclusion applies also to any other excitation
  except for the $K^{\pi}=0^+$ states.  Furthermore, the
  difference between VP and SP is found small (with exception of
  peak height in the toroidal mode). In the low-energy
    isovector dipole (pygmy) and isoscalar toroidal modes, the branch
  $K^{\pi}=1^-$ is shown to dominate over $K^{\pi}=0^-$ one in
    the range of excitation energy $E <$ 8--10 MeV. The effect
  becomes impressive for the toroidal resonance whose low-energy part
  is concentrated in a high peak of almost pure $K^{\pi}=1^-$
  nature. This peculiarity may be used as a fingerprint of the
  toroidal mode in future experiments. The interplay between pygmy, toroidal
  and compression resonances is discussed, the interpretation of the observed
  isoscalar giant dipole resonance is partly revised.
\end{abstract}

\maketitle
\section{Introduction}
\label{intro}

Pairing is known to play an important role in low-energy nuclear
excitations \cite{SSS,Ri80}. For the ground state, it is usually
treated within the Bardeen-Cooper-Shrieffer (BCS) scheme or
Hartree-Fock-Bogoliubov (HFB) method, see
e.g. \cite{SSS,Ri80,Do84,Do96,Re97,Be00,Ta04}. Nuclear
  excitations are described with the Quasiparticle Random Phase
Approximation (QRPA) where the pairing-induced particle-particle (pp)
channel is added to the particle-hole (ph) residual interaction
\cite{SSS,Ri80,Te05,Li08}. During last decades, both static and
dynamical pairing aspects were thoroughly explored in schematic and
self-consistent models, see
e.g. \cite{SSS,Do84,Do96,Re97,Be00,Ta04,Te05,Li08,Iud05,Se08,Te10,Te11g}. The
pp-channel was found important in electric modes with positive
  parity (monopole, quadrupole) where it generates pairing vibrations,
see e.g. \cite{SSS,Ri80,Li08,Iud05}.  In deformed axial nuclei, the
pp-channel is important in states with $K^{\pi}=0^+$ where $K$ is the
projection of the angular momentum \cite{SSS,Iud05,Te10,Te11g}.

In spite of extensive studies, some pairing effects in nuclear excitations
still deserve further exploration, especially in deformed nuclei. For
example, it is yet unclear if we indeed need the pp-channel in the
self-consistent description of excitations with $K^{\pi} \ne
0^+$. These excitations include all negative-parity states (dipole,
octupole, ...) as well as positive-parity states with $K>0$.
Clarification of this point is important because the pp-channel
considerably complicates QRPA calculations.
A further point of interest is to check the influence of
surface pairing (SP) as compared to volume pairing (VP).

These two problems are addressed in the present study for the case
of the standard monopole $\delta$-force  pairing. The calculations
are performed within the self-consistent QRPA method \cite{Re16}
with the Skyrme force SLy6 \cite{Cha97}. As typical examples we consider
$^{152,154,156}$Sm which are well deformed having still axially
symmetric shape. In a first step, we inspect the lowest
$\gamma$-vibrational states with $K^{\pi}=2^+$.  The previous
systematic self-consistent explorations of $\gamma$-vibrational states
were performed with \cite{Te11g} and without \cite{Ne16g} the
pp-channel.  We will show that the effect of pp-channel is very small
in these modes. At least it is much weaker than the difference
between the results obtained with VP and SP.

In a second step, we inspect exotic parts of the dipole
  excitation spectrum which attract high attention presently (see,
  e.g., \cite{Pa07,Kv11,Re13}), namely low-lying isovector strength
  also coined pygmy dipole resonance (PDR), toroidal dipole resonance
  (TDR), and compression dipole resonance (CDR).  These modes are
located, fully or partly, at rather low energies and so may be
affected by pairing.  Here again we have found the pp-channel
negligible. Altogether, our analysis suggests that the
pp-channel can be ignored in QRPA calculations for {\it all
}excitations with $K^{\pi} \ne 0^+$.

The next point of our study concerns deformation effects in dipole
excitations.  We analyze the ordinary isovector giant dipole resonance
(GDR), PDR, and isoscalar TDR and CDR. The TDR and CDR are often
considered as low- and high-energy parts of the isoscalar
giant dipole resonance (ISGDR) \cite{Pa07,Har}.
In axial nuclei, all these resonances exhibit a
deformation splitting into $K=0$ and $K=1$ branches. However,
the features of these branches depend very much on the
actual type of resonance. For example, our previous studies have
shown that the main low-energy peak of the TDR is, somewhat
  surprisingly, dominated by $K=1$ strength instead of $K=0$ one as
would have expected \cite{Kv13Sm,Kv14Yb,Ne17}.  In fact, the TDR
produces the opposite sequence of $K$-branches as compared to
the GDR. Since the TDR shares the same energy region as the PDR and
is, in fact, the source of the PDR \cite{Re13}, this anomaly
becomes especially interesting because it can be related to the
deformation splitting of the PDR.  In this connection, we propose here
a further comparative analysis of the deformation splitting of
various dipole modes with the main accent to the low-energy region
embracing PDR and TDR.

The paper is organized as follows. In Sec. 2, the calculation scheme
is outlined paying particular attention to the treatment of
pairing. The main factors determining the impact of the pp-channel are
discussed comparing analytically the pp- and ph-channels.  In
Sec. 3, results of the QRPA calculations for $K^{\pi}=2^+$ states and
dipole modes are presented and discussed. Some experimental perspectives
are commented. In Sec. 4, conclusions are
given. In Appendix A, the Skyrme functional used in this paper is
sketched. In Appendix B, the actual HF+BCS scheme is outlined. In
Appendix C, the basic QRPA equations with pp-channel are presented.

\section{Calculation scheme and theoretical background}
\label{sec-1-theor}
\subsection{Method and numerical details}
\label{subsec21}

The calculations are performed within QRPA
\cite{Ri80} based on the Skyrme energy functional \cite{Skyrme,Vau,Be03}
\begin{equation}
\mathcal{E}(\rho,\tau, {\bf J}, \vec{j}, \vec{s}, \vec{T}, \tilde{\rho})
= \mathcal{E}_{\rm kin} + \mathcal{E}_{\rm Sk}
+ \mathcal{E}_{\rm Coul} + \mathcal{E}_{\rm pair}
\label{Efunc}
\end{equation}
including kinetic, Skyrme, Coulomb and pairing parts.
The Skyrme part $\mathcal{E}_{\rm{Sk}}$ depends on the following local
densities and currents: density $\rho(\vec{r})$, kinetic-energy
density $\tau(\vec{r})$, spin-orbit density $\mathbf{J}(\vec{r})$,
current $\vec{j}(\vec{r})$, spin density $\vec{s}(\vec{r})$, and spin
kinetic-energy density $\vec{T}(\vec{r})$. The Coulomb exchange term
is treated in Slater approximation.  The pairing functional is
derived from monopole contact $\delta$-force interaction \cite{Be00}.
This functional
depends on the pairing
density $\tilde{\rho}(\mathbf{r})$ and, for the surface pairing, also
on the normal density $\rho(\mathbf{r})$. More details on the
functional (\ref{Efunc}) can be found in the next subsection and in
Appendix A.

The mean-field Hamiltonian and the QRPA residual interaction are
determined through the first and second functional derivatives of
  the total energy (\ref{Efunc}).  The approach is fully
self-consistent since: i) both the mean field and residual interaction
are obtained from the same Skyrme functional, ii) time-even and
time-odd densities are involved, iii) the residual interaction
includes all the terms of the initial Skyrme functional as well as the
Coulomb direct and exchange terms, iv) both ph- and pp-channels in
the residual interaction are taken into account.
Details of our mean field and QRPA schemes are given in
Appendices B and C.  The present version of the Skyrme QRPA scheme is
implemented in the two-dimensional (2D) QRPA code \cite{Re16}.

The calculations are performed with the Skyrme force SLy6
\cite{Cha97} which was successfully used earlier for a systematic
exploration of the GDR in rare-earth, actinide and superheavy nuclei
\cite{Kl08}.
The QRPA code employs a 2D mesh in cylindrical coordinates.  The
calculation box extends over three times the nuclear radius. The
mesh size is 0.4 fm. The single-particle spectrum taken into
  account embraces all levels from the bottom of the potential well
up to +30 MeV.  The (axial) equilibrium quadrupole deformation is
determined by minimization of the total nuclear energy. This
  yields the deformation parameters $\beta$ = 0.311, 0.339, and 0.361
for $^{152,154,156}$Sm, respectively, which are sufficiently
  close to the experimental values  $\beta_{\rm exp}$ = 0.308 for
$^{152}$Sm and $\beta_{\rm exp}$ = 0.339 for $^{154}$Sm
\cite{bnl_exp}.

Pairing in the ground state is treated with the HF+BCS scheme,
see next subsection. To cope with the principle divergency of
short-range forces, we implement in the HF+BCS equations and pairing
transition densities a soft, energy-dependent cutoff factor
\cite{Be00}
\begin{equation}
 f^q_i
 =
 \frac{1}
      {1+\exp\Big[\frac{e^q_i-\lambda_{q}-\Delta E_q}{\eta_{q}}\Big]}
 \label{f-cut}
\end{equation}
where $q=\{n,p\}$, $i$ denotes single-particle states, $e^q_i$ is the
single-particle energy, and $\lambda_q$ is the chemical potential. The
cutoff parameter $\Delta E_q$ and width $\eta_{q}=\Delta E_{q}/10$ are
chosen self-adjusting to the actual level density in the vicinity of
the Fermi energy \cite{Be00}. Here $\Delta E_q$ lies around 6 MeV.  In
the pp-channel, two-quasiparticle (2qp) states $(i,j)$ are
  limited to those with $\sqrt{f^q_if^q_j} > 10^{-4}$ to remove
high-energy particle states.  More information on the pairing
formalism is given in the next subsection.

To minimize the calculational expense in ph-channel, we employ
only states with $|u_iv_j+v_iu_j| > 10^{-4}$, where $u_i,v_i$ are
coefficients of the special Bogoliubov transformation. This cut-off
still leaves the 2qp configuration space large enough to exhaust
the energy-weighted sum rules (EWSR).
Note also that a large configuration space in ph-channel is crucial
for a successful description of low-energy vibrational excitations
\cite{Ne16g} and extraction of the spurious admixtures
\cite{Re16,Rei92d}. For example, in calculations of
$\gamma$-vibrational modes in $^{154}$Sm, we deal with $\approx
6300$ proton  and $\approx 12500$ neutron pairs and exhaust the
isoscalar quadrupole $\mathrm{EWSR}=(\hbar e)^2/(8\pi m)\cdot 50A\langle
r^2\rangle_A$ by $\approx$ 97 $\%$. The Thomas-Reiche-Kuhn sum rule
for dipole excitations is exhausted in the interval 2--45 MeV by
$\approx$ 94 $\%$.

\subsection{Treatment of pairing}
\label{subsec22}

The pairing part in the functional (\ref{1}) reads
\begin{equation}
\mathcal{E}_{\rm pair} = \frac{1}{4} \: \sum_{q=n,p} V_q \: \int \mathrm{d}\vec{r} \:
|\tilde{\rho}_q (\vec{r})|^2 \:  G_q(\vec{r})
\;.
\label{Epair}
\end{equation}
It is motivated by the zero-range pairing interaction
\cite{Be00}
\begin{equation}
  V^q_{\rm pair}(\vec{r},\vec{r}') = G_q(\vec{r})\delta(\vec{r}-\vec{r}')
\label{Vpair}
\end{equation}
with
\begin{equation}
 G_q(\vec{r})= V_q \Big[ 1 - \eta \:
\Big(\frac{\rho(\vec{r})}{\rho_0}\Big)^{\!\gamma}\Big] .
\label{Gpair}
\end{equation}
Here $\rho(\vec{r})=\rho_p(\vec{r})+\rho_n(\vec{r})$ is the sum of
normal proton and neutron densities, $\tilde\rho_q(\vec{r})$ are the
pairing densities, $V_q$ are pairing strength constants, and $\rho_0$=0.16
${\rm fm}^{-1}$ is the saturation density of symmetric nuclear
matter. The coefficient $\gamma$ regulates the density
dependence. Following common practice, we fix here $\gamma$=1.
We obtain so-called volume pairing (VP) for $\eta$=0  and the
(density-dependent) surface pairing (SP) for $\eta$=1. Of
course, $\eta$ may be varied freely and take values in between 0 and 1,
when optimized properly \cite{Klu09a}. We consider here VP and SP
as limiting cases to explore the sensitivity to the pairing model.

Pairing is treated within the HF+BCS scheme described in Appendix
B.  In this scheme, the nucleon and pairing densities in axial nuclei
read \cite{Be03}
\begin{eqnarray}
\label{rho}
\rho_q(\vec{r})&=& 2 \sum_{i \epsilon q} v_i^2 |\psi_i (\vec{r})|^2 ,
\\
\tilde\rho_q(\vec{r}) &=& -2 \sum_{i \epsilon q} f_i^q v_i u_i |\psi_i (\vec{r})|^2
\label{rho_pair}
\end{eqnarray}
where $\psi_i (\vec r)$ are single-particle wave functions and $f^q_i$
is the energy-dependent cut-off weight (\ref{f-cut}).

Pairing induces the pp-channel in the QRPA residual interaction, see
Appendix C. Its contributions to the QRPA matrices A and B are given
by expressions (\ref{C.1})-(\ref{C.4}). For $\gamma$=1, these
contributions read:
\begin{eqnarray}
A^{\rm pair}_{qq'}
(i{\bar j}, i'{\bar j}')
&=&
\frac{1}{2} \delta_{q q'} V_q  \int \mathrm{d}\vec{r}
\Big[ 1 - \eta \frac{\rho({\vec r})}{\rho_0} \Big]
\delta\tilde{\rho}^q_{i{\bar j}}(\vec{r})
\delta\tilde{\rho}^{q'}_{i'{\bar j}'}(\vec{r})
\nonumber
\\
&-& \frac{\eta}{2 \rho_0} V_q
\int \mathrm{d}\vec{r} \tilde{\rho}_q(\vec{r})
\delta\tilde{\rho}^q_{i\bar{j}}(\vec{r})
\delta{\rho}^{q'}_{i'\bar{j}'}(\vec{r})
\nonumber\\
&-& \frac{\eta}{2 \rho_0}  V_{q'}
\int \mathrm{d}\vec{r} \tilde{\rho}_{q'}(\vec{r})
\delta\rho^q_{i\bar{j}}(\vec{r})
\delta\tilde{\rho}^{q'}_{i'\bar{j}'}(\vec{r})
\label{RPA_App}
\end{eqnarray}
with
\begin{equation}
B^{\rm pair}_{i {\bar j} i' {\bar j}'} = -A^{\rm pair}_{i {\bar j}\: i' {\bar j}'} .
\label{B}
\end{equation}
Here we assume that $i{\bar j} \in q, \: i'{\bar j}' \in q'$. Furthermore
\begin{equation}
\delta\rho^q_{i{\bar j}}(\vec{r})=
(u_i v_j + u_j v_i) \psi^*_i(\vec{r})\psi_j(\vec{r})
\label{TDn}
\end{equation}
\begin{equation}
\delta\tilde{\rho}^q_{i{\bar j}}(\vec{r})= \sqrt{f^q_i f^q_j}\:
(u_i u_j - v_j v_i) \psi^*_i(\vec{r})\psi_j(\vec{r})
\label{TDp}
\end{equation}
are 2qp nucleon and pairing transition densities.

The contributions (\ref{RPA_App})-(\ref{B}) constitute the
pairing-induced pp-channel.
The first term in (\ref{RPA_App}) exists for both VP and SP. It is
diagonal in $q$ since we do not consider a proton-neutron pairing.
The next two terms which mix nucleon and pairing density are
caused by the density-dependence of SP and so exist only in SP
case. In the present study we consider only quadrupole states $K^{\pi}
= 2^+$ and dipole states. Then the transition densities (\ref{TDn})
and (\ref{TDp}) include only non-diagonal 2qp configurations
with $i \ne j$.

It is instructive to compare the contributions to QRPA matrices from
ph- and pp-channels.  The ph-contribution to the A-matrix from the
Skyrme functional is dominated by \cite{Ne06}
\begin{equation}
\frac{\delta\mathcal{E}_{\rm Sk}}{\delta\rho^q \delta\rho^{q'}}= b_0 -b_0'\delta_{qq'}
\end{equation}
where $b_0$ and $b_0'$ are coefficients of the Skyrme interaction
given in the Appendix A.  This results in
\begin{equation}
A^{\rm ph}_{qq'}
(i{\bar j}, i'{\bar j}')
=
(b_0 -b_0'\delta_{qq'}) \int d\vec{r}
\delta\rho^q_{i{\bar j}}(\vec{r})
\delta\rho^{q'}_{i'{\bar j}'}(\vec{r})
\end{equation}
to be compared for $q=q'$ with the pp-contribution (\ref{RPA_App}).
For states $i$ and $j$ near the Fermi level, the transition
densities (\ref{TDn}) and (\ref{TDp}) are of the same order of magnitude.
Then the ratio between ph and pp contributions
is mainly determined by the ratio of the channel strengths
\begin{equation}
  R_q = \frac{2(b_0 -b_0')}{V_q} .
\end{equation}
For SLy6, $b_0=-3502.3$ MeV, $b_0'=-3285.3$ MeV. Further, $V_p=-298.8$
MeV fm$^3$, $V_n=-288.5$ MeV fm$^3$ for VP and
$V_p=-1053.12$ MeV fm$^3$, $V_n=-864.2$ MeV fm$^3$ for SP \cite{Fle04}.
Then the ratios are $R_n=1.50$,  $R_p=1.45$ for VP and $R_n=0.50$,
$R_p=0.41$ for SP. In average, $R_q \sim 1$, i.e. Skyrme
amplitudes in ph-channel and pairing amplitudes in pp-channel are
similar (though, for SP, one may expect somewhat larger effect
of pp-channel than for VP).

The observations that $\rho^q_{ij} \sim \tilde\rho^q_{ij}$ and $R_q
\sim $1 do not yet suffice to conclude that ph- and pp-channels
are of the same order of magnitude. We should also take into account
that in ph-channel the main interaction is the proton-neutron one
(with the amplitude $b_0$).  This interaction is much stronger than
its proton-proton and neutron-neutron counterparts characterized by
the amplitudes $b_0-b_0'$. Indeed, for SLy6,
$b_0/(b_0-b_0')=16.1$. Pairing, on the other hand, has no
neutron-proton interaction at all. Besides, in the pairing pp-channel,
the contribution of 2qp configurations with high-energy states is
suppressed by the cut-off (\ref{f-cut}).  So in general one may expect
that the ph-channel is much stronger than the pp-channel.

Note that the above qualitative analysis mainly concerns states with
$K^{\pi} \ne 0^+$. For $K^{\pi} = 0^+$ states, the situation becomes
much more complicated. Here the diagonal 2qp configurations $i
\bar{i}$, resulting in the pairing vibrations, come into play. Thus
the pp-channel, even being still
weaker than its ph-counterpart, becomes important for low-energy
excitations \cite{SSS,Ri80}.

\subsection{QRPA strength functions}
\label{subsec23}

The energy-weighted strength function for electric transitions
of multipolarity $\lambda\mu$ between QRPA ground state
$|{\rm gr}\rangle$ and excited state $|\nu\rangle$ has the form
\begin{equation}
S_X(E\lambda\mu, T; E) = \sum_{\nu} E_{\nu} \: \big| \langle \nu|\:
\hat{M}^X_{\lambda \mu}(T) \: |{\rm gr} \rangle \big|^2 \: \xi_{\Delta}(E-E_{\nu})
\label{13}
\end{equation}
where $E$ is the excitation energy, $\hat{M}^X_{\lambda \mu}(T)$
is the transition operator, $X=\{\rm el, com, tor\}$ marks the type
of the excitation (isovector E1(T=1), compression isoscalar E1(T=0),
toroidal isoscalar E1(T=0)), $E_{\nu}$ is the energy of the QRPA state.
For excited states $K^{\pi}_{\nu}$, we assume $K=\mu$, $\pi=(-1)^{\lambda}$.
The total strength functions read
\begin{equation}
S_X(E\lambda, T; E) = \sum_{\mu \ge 0 } (2-\delta_{\mu,0}) S_X(E\lambda\mu, T; E) .
\label{13a}
\end{equation}
The strengths are weighted by the Lorentz function
\begin{equation}
\xi_{\Delta}(E-E_{\nu}) = \frac{1}{2 \pi}
\frac{\Delta (E)}{(E-E_{\nu})^2 + [\Delta(E)/2]^2}
\label{17}
\end{equation}
with the energy-dependent folding \cite{Kv13Sm}
\begin{equation}
\Delta(E) =
\left\{
\begin{array}{ll}
 \Delta_0 & \:{\rm for} \:\:\: E\leq E_0, \\
 \Delta_0 + a\:  (E-E_0) &\: {\rm for} \:\:\: E> E_0.
\end{array} \right.
\label{18}
\end{equation}
The weight $\xi_{\Delta}(E-E_{\nu})$ is used to simulate escape width
and coupling to the complex configurations. Since these two effects
grow with increasing the excitation energy, we employ an energy
  dependent folding width $\Delta(E)$.  The parameters $\Delta_0$=0.1
MeV, $E_0$=8.3 MeV and $a=$0.45 are chosen so as to reproduce the
smoothness of the experimental photoabsorprion  in
$^{154}$Sm.  The value $E_0$ roughly corresponds to the neutron
and proton separation energies in $^{154}$Sm: $S_n$=8.0 MeV and
$S_p$=9.1 MeV \cite{bnl_exp}.

The operator for the electric $E1\mu (T=1)$
transition reads
\begin{equation}
\hat{M}^{el}_{1\mu}(T=1) = e \sum_{i=1}^{A}
e^{\rm eff}_i \: r_i  Y_{1\mu}(\Omega_i)
\label{14}
\end{equation}
where $Y_{1\mu}(\Omega_i)$ is the spherical harmonic and
$e^{\rm eff}_i$ are effective charges
equal to $N/A$ for protons and $-Z/A$ for neutrons.
The operators for the compression and toroidal $E1\mu (T=0)$
transitions are \cite{Kv11}
\begin{equation}
\label{Tcom}
\hat{M}^{\rm com}_{1\mu} (T=0)
= \frac{e}{10} \:
\sum_{i=1}^{A} [ r_i^3
-\frac{5}{3}\langle r^2 \rangle_0 r_i ]
Y_{1 \mu}(\Omega_i) ,
\end{equation}
\begin{eqnarray}
&&
\hat{M}^{\rm tor}_{1\mu}(T=0) =
 - \frac{1}{2c\sqrt{3}}
\int d\vec{r} \:
\hat{{\bf j}}_{\rm nuc}(\vec{r}-\vec{r}_i)
\nonumber \\
&&
\cdot
\big[
\frac{\sqrt{2}}{5} r^2_i {\bf Y}_{12\mu}(\Omega_i)
+ (r^2_i - \langle r^2 \rangle_0)
{\bf Y}_{10\mu}(\Omega_i)
\big]
 \label{15}
\end{eqnarray}
where ${\bf Y}_{10\mu}(\Omega_i)$, ${\bf Y}_{12\mu}(\Omega_i)$ are
vector spherical harmonics;
$
\hat{\vec{j}}_{\rm nuc}(\vec{r}-\vec{r}_i)
= -ie\hbar/(2m) \sum_{i=1}^A [\delta(\vec{r}-\vec{r}_i)\vec{\nabla}_i
+\vec{\nabla}_i\delta(\vec{r}-\vec{r}_i)]
$ is
operator of the isoscalar convection nuclear current.  The terms with the
squared ground-state radius $\langle r^2 \rangle_0$ represent the
center-of-mass-corrections. Note that the compression operator (\ref{Tcom}) is
also the transition (probe) operator for ISGDR \cite{Har}.

The photoabsorption \cite{Ri80} is related to the strength function (\ref{13a}) as
\begin{equation}
\sigma (E) \: [{\rm fm}^2] \: \approx 0.402 \:  S_{\rm el}(E1, T=1;\:E) .
\label{16}
\end{equation}

\section{Results and discussion}
\label{sec-3}
\subsection{$\gamma$-vibrational states}
\label{subsec31}

\begin{figure}
\resizebox{0.35\textwidth}{!}{%
  \includegraphics{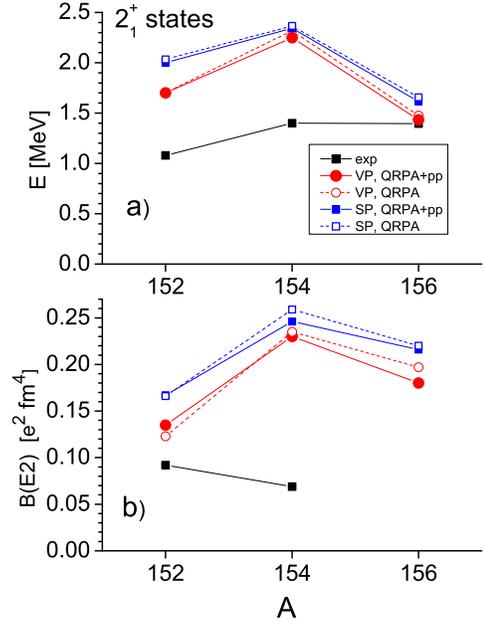}}
\vspace{0.3cm}
\caption{Energies (upper panel) and B(E2) values (lower panel) for
the lowest gamma-vibrational states $K^{\pi}_i=2^+_1$ in
$^{152,154,156}$Sm, calculated
with (full symbols) and without (open symbols) the pp-channel. Both
cases of volume (VP) and surface (SP) pairing are considered. The
experimental data are taken from \cite{bnl_exp}.}
\label{fig:1}
\end{figure}

\begin{figure}
\resizebox{0.47\textwidth}{!}{%
  \includegraphics{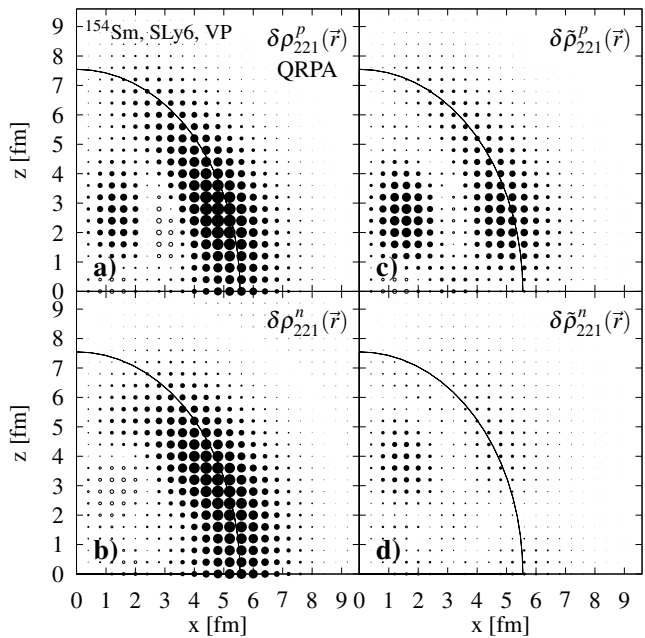}}
\caption{Nucleon (left) and pairing (right) proton and neutron QRPA
  transition densities for the lowest gamma-vibrational $K^{\pi}_i=2^+_1$
  -state in $^{154}$Sm, calculated with VP.  The densities are plotted
  in the x-z plane for $x,z \ge 0$. The filled (open) circles represent
  positive (negative) values. Magnitudes of the densities are depicted
  by the circle size (in arbitrary units).  The nuclear surface is
  marked by the solid line.}
\label{fig:2}       
\end{figure}

\begin{figure}
\resizebox{0.47\textwidth}{!}{%
  \includegraphics{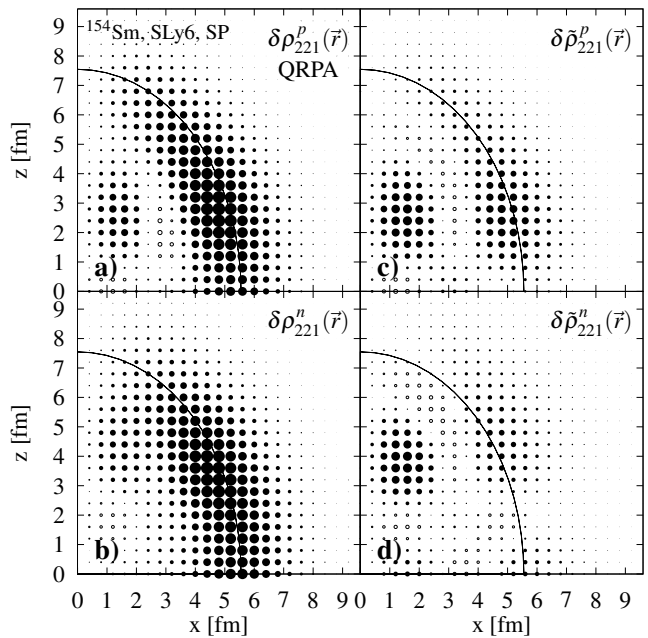}}
\caption{The same as in Fig. 2 but for SP.}
\label{fig:3}       
\end{figure}

\begin{figure}
\resizebox{0.47\textwidth}{!}{%
  \includegraphics{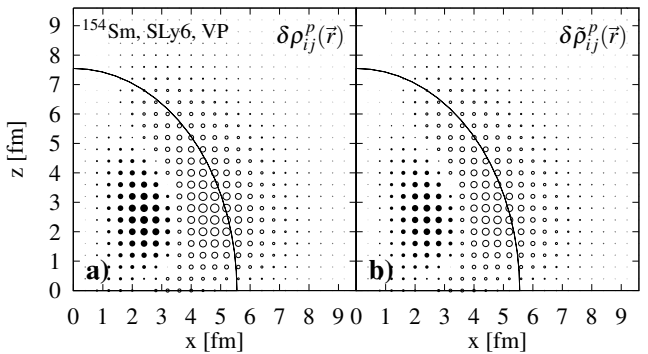}}
 \vspace{0.8cm}
\caption{Nucleon TD (\ref{TDn}) and pairing TD (\ref{TDp}) for the
 lowest 2qp proton excitation $ij=pp[413\downarrow - 411\downarrow]$ in
 $^{154}$Sm, calculated for VP.  See caption to Fig. 2 for more
 details.}
\label{fig:4}       %
\end{figure}

Figs. \ref{fig:1}-\ref{fig:4} collect results of our calculations for
$\gamma$-vibrational states in $^{152,154,156}$Sm.  Fig. \ref{fig:1}
shows energies and $B(E2;\: gr \rightarrow I^{\pi}K=2^+2)$ values
computed with and without the pp-channel for VP and SP cases.
The reduced transition probabilities are obtained with the
effective charges $e^{\rm eff} =1$ for protons and $e^{\rm eff} =0$
for neutrons.  Fig. \ref{fig:1} indicates that the present QRPA
description is not fully satisfying: the calculations
essentially overestimate the experimental energies in $^{152,154}$Sm
and $B(E2)$ in $^{154}$Sm. Note that such significant overestimation
of energies of $\gamma$-vibrational states in $^{152,154}$Sm has been
already observed in previous self-consistent Skyrme QRPA
calculations \cite{Te11g,Ne16g}. It was shown \cite{Ne16g}
that this problem can be partly solved by taking into account the
pairing blocking effect and using a more appropriate Skyrme
parametrization. Besides that, the coupling with complex (two-phonon)
configurations could be here important \cite{SSS}.

As seen from Fig. \ref{fig:1}, VP and SP results are rather similar
though VP performance is somewhat better. What is remarkable, for both
VP and SP, the impact of the pp-channel is very small, which actually
means that the pp-channel is much weaker than the ph-one (at least its
impact is much smaller than the difference between VP and SP results).

This conclusion can also be drawn from Fig. \ref{fig:2} which
shows the pairing and nucleon transition densities (TD) for the
$\gamma$-vibrational state ($\lambda\mu\nu$ = 221) in $^{154}$Sm,
using VP. We see that, at the nuclear surface, nucleon TD
$\delta\rho^q_{221}$ are much larger than pairing TD
$\delta\tilde{\rho}^q_{221}$, especially for neutrons.  The
same takes place in Fig. \ref{fig:3} for the case of SP.

The next question is: what is the origin of the difference in the
nucleon and pairing one-phonon TD?  To clarify this point, we should
  numerically analyze the main ingredients of the contributions
(\ref{RPA_App})--(\ref{B}) of the pp-channel to the RPA matrices.
First of all, let's consider nucleon (\ref{TDn}) and pairing
(\ref{TDp}) TD for the main 2qp component of the $\gamma$-vibrational
state. This is the proton $F,F+3$ configuration
$pp[413\downarrow-411\downarrow]$ (here we use asymptotic Nilsson
quantum numbers, $F$ denotes the Fermi level).  The TD for this
configuration are shown in Fig. \ref{fig:4} for the case of VP.  It is seen
that the nucleon and pairing TD are very similar. This is not surprising
since both TD have the same coordinate dependence and deviate only by
the pairing factors which are of the same order of magnitude for the
states near the Fermi level. As shown in Sec.  \ref{subsec22}, the
strength coefficients of the pp- and ph-channels are also similar.  So
indeed, following the discussion in Sec. \ref{subsec22}, the weakness
of the pp-channel is perhaps mainly caused by the absence of the
neutron-proton interaction (which plays a dominant role in
ph-channel).

The conclusion that the pp-channel plays a negligible role
can, in principle, be generalized to arbitrary states with $K^{\pi}
\ne 0^+$ in even-even nuclei, e.g. to all states of negative
parity. This allows to ignore the pp-channel from the onset and
thus considerably decrease the numerical expense of QRPA.

\subsection{Dipole excitations}
\label{subsec32}

The previous subsection concluded that pp-channel has negligible
impact for low-energy $K^{\pi} \ne 0^+$ states. This should be even
more the case for excitations with higher energies, e.g. for the
various dipole resonances. For these modes, the pairing gap $\Delta_q
\sim$ 2 MeV is much smaller than the resonance energies and so the
impact of pairing should be even weaker than for low-energy excitations.
Note that low-energy excitations are composed from 2qp pairs whose
single-particle states $\{i,j\}$ are placed near the Fermi level and
so are strongly affected by pairing.  Instead, in high-energy
excitations, at most one single-particle state from the 2qp pair can
originate from the Fermi level region. Thus in general high-energy 2qp
excitations are much less affected by pairing than low-energy ones and
so, in principle, should be only weakly affected by the
pairing-induced pp-channel.  For the same reason, the difference
between VP and SP in the resonance region is also expected to be
negligible.  This will be confirmed below by our QRPA calculations for
GDR, PDR, TDR and CDR in $^{154}$Sm. In this subsection, we will
consider peculiarities of the deformation splitting of the dipole
resonances in this nucleus.  The main attention will be paid to a
comparison of their $K=0$ and $K=1$ branches in the low-lying energy
interval 3--9 MeV, embracing PDR and TDR. In what follows, it is
assumed that $K=1$ branch includes both $K=\pm $1 contributions.

Use of a large configuration space allows to keep the spurious
center-of-mass mode at low energies $<2$ MeV. The elimination
of this mode from the dipole strength functions is
additionally ensured by using appropriate effective charges and
correction terms, see Sec. 2.3. In the following, we discuss
dipole strength functions only for the excitation energy $E > $ 3
MeV, which is at the safe side with respect to the center-of-mass
  mode.
\begin{figure}
\resizebox{0.48\textwidth}{!}{%
\includegraphics{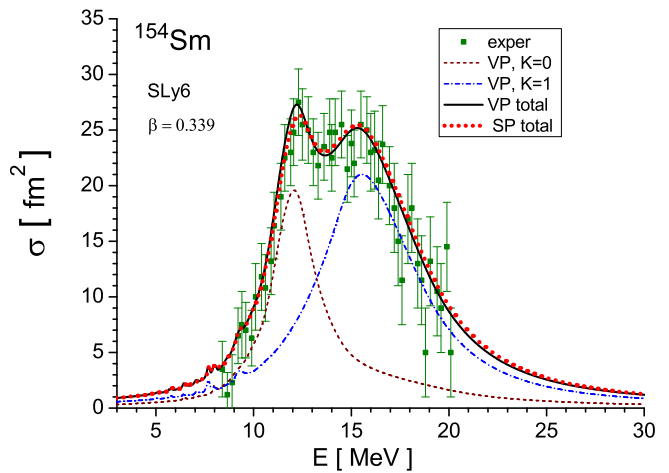}}
\caption{Calculated and experimental \cite{GDRexp}  photoabsorption cross-sections
in $^{154}$Sm. The total strengths are calculated with the pp-channel for VP
and SP. For VP, $K=0$ and $K=1$ components are shown.}
\label{fig:5}
\end{figure}

Fig. \ref{fig:5} compares the QRPA photoabsorption
strength with the experimental data \cite{GDRexp}. We see an excellent
agreement between the theory and experiment. As said
above, an appropriate modeling of the GDR shape is
achieved by energy-dependent averaging described in Sec. 2.3.
However, a correct description of peak position and, in a large extent, of fragmentation
widths is achieved by QRPA with the parametrization SLy6. Altogether,
the obtained agreement validates  our approach for the further studies.
Fig. \ref{fig:5} shows that VP and SP give almost identical results.
The pp-channel does not have any effect (not shown since curves
with and without the pp-channel almost coincide). The GDR exhibits
deformation splitting into $K=0$ and $K=1$ branches.  The sequence of
branches is typical for GDR in prolate nuclei: $K=0$ branch lies below
of $K=1$ one. Though $K=0$ branch involves about twice less strength
than $K=1$ one, its peak height is a bit larger. This is
explained by a stronger fragmentation of the $K=1$ branch, which
broadens the distribution.
\begin{figure}
\resizebox{0.48\textwidth}{!}{%
\includegraphics{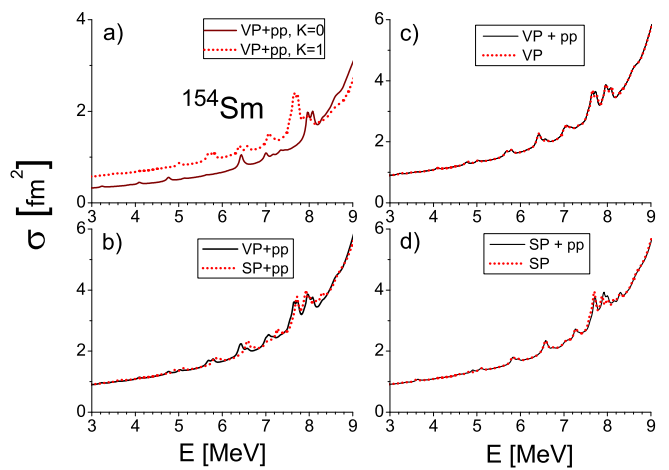}}
\vspace{0.3cm}
\caption{The calculated photoabsorption in the region of the pygmy
  resonance in $^{154}$Sm.  (a) K=0 and K=1 strengths with VP and
  pp-channel: b) total strengths (sums of $K=0$ and $K=1$ contributions)
  for VP and SP with pp-channel, c) total strengths for VP with and without pp-channel,
  d) the same as c) but for SP.}
\label{fig:6}
\end{figure}

Fig. \ref{fig:6} shows the calculated photoabsorption in the PDR
region.  Following plots c)-d), the influence of the pp-channel is
very weak for VP and SP (though for SP there are some small
differences for the structures around 8 MeV).  In the plot b),
results from VP and SP deviate only in detail.  Plot a) shows
that $K=0$ strength exceeds $K=1$ strength at $E \ge$ 8 MeV but becomes
smaller at $E \le$ 8 MeV. A similar effect was found for Mo isotopes
\cite{Mo}.  This is the result of the competition between GDR tails
from $K=0$ and $K=1$ branches.
\begin{figure}
\resizebox{0.45\textwidth}{!}{%
\vspace{0.8cm}
\includegraphics{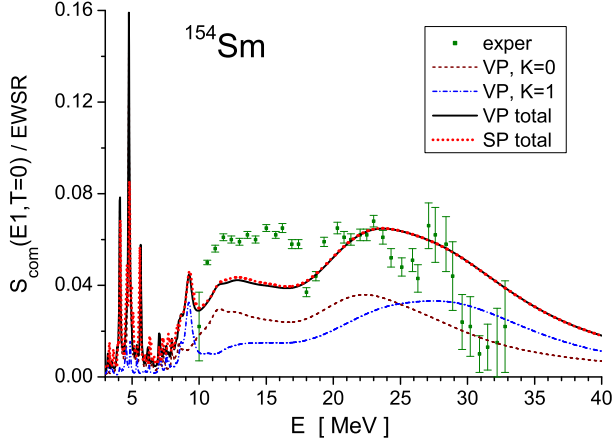}}
\caption{The compression isoscalar E1 strength (ISGDR) in $^{154}$Sm,
calculated with pp-channel for VP and SP cases. For VP, the $K=0$ and $K=1$
components are shown. The calculated strength is compared with the
experimental data \cite{CDRexp}. Calculated and experimental strengths are given
as fractions of the corresponding EWSR(E1,T=0). See text for more detail.}
\label{fig:7}
\end{figure}
\begin{figure}
\resizebox{0.48\textwidth}{!}{%
\includegraphics{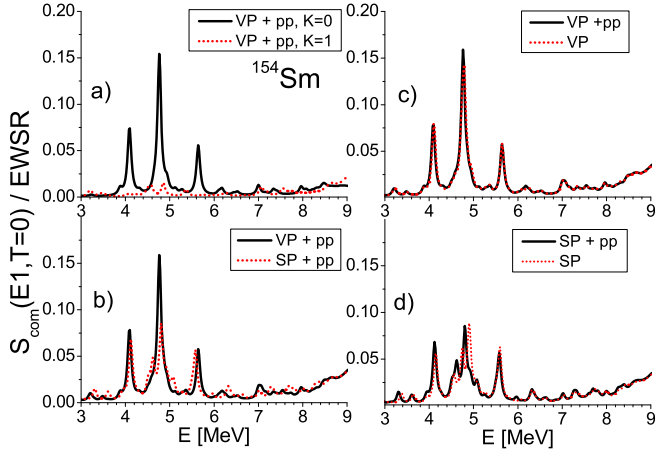}}
\vspace{0.3cm} \caption{The calculated low-energy part of ISGDR in $^{154}$Sm.
(a) VP with pp channel: K=0 and K=1 components, b) VP and SP with pp-channel, c)
VP with and without pp-channel, d) SP with and without pp-channel.} \label{fig:8}
\end{figure}
\begin{figure}
\resizebox{0.45\textwidth}{!}{%
  \includegraphics{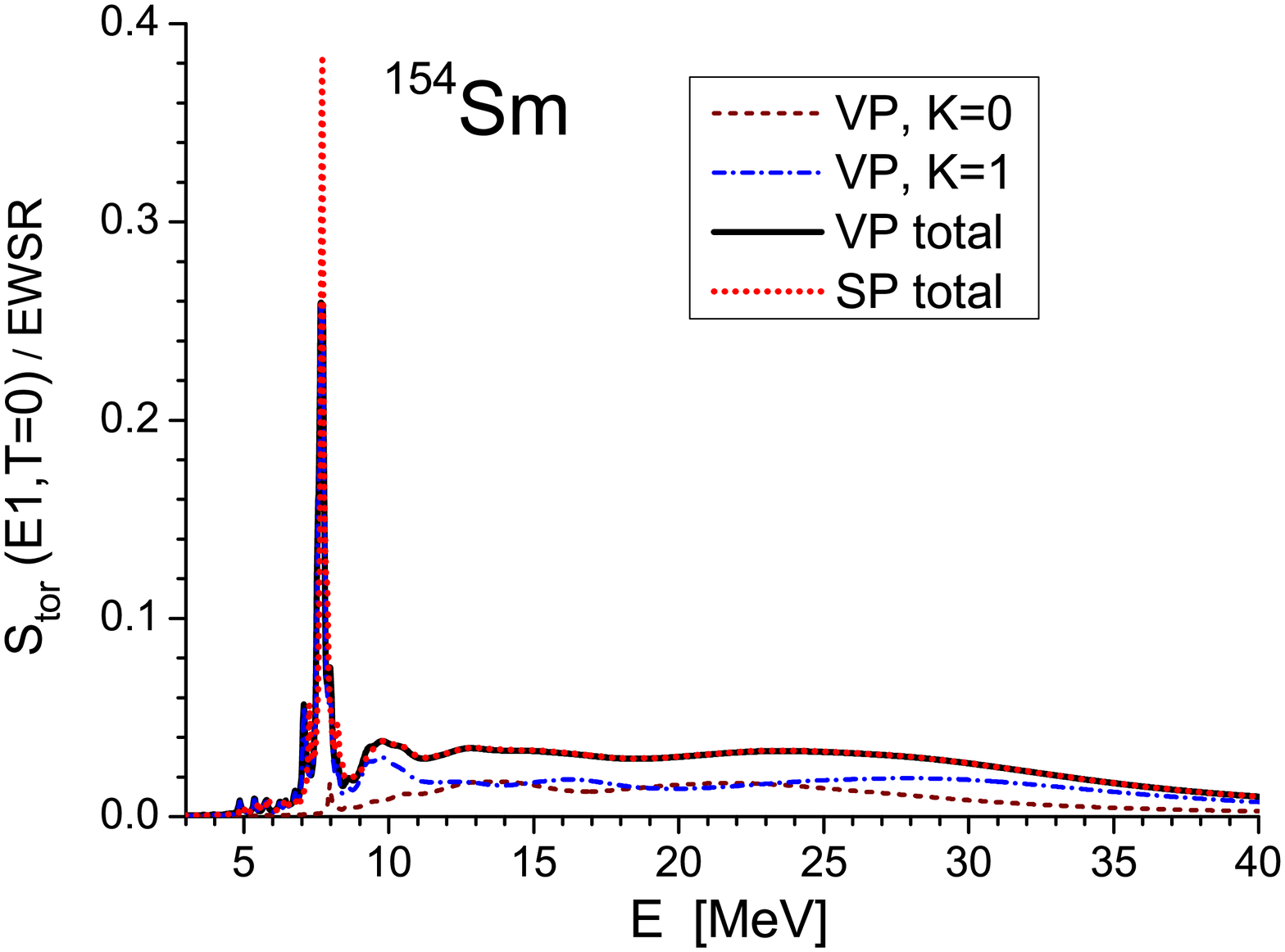}}
\vspace{0.3cm}
\caption{The isoscalar toroidal E1 strength function in $^{154}$Sm,
  calculated for volume (VP) and surface (SP) pairing. In both cases,
  the pp-channel is included. For VP, the $K=0$ and $K=1$ components
  are shown.}
\label{fig:9}
\end{figure}
\begin{figure}
\resizebox{0.48\textwidth}{!}{%
\includegraphics{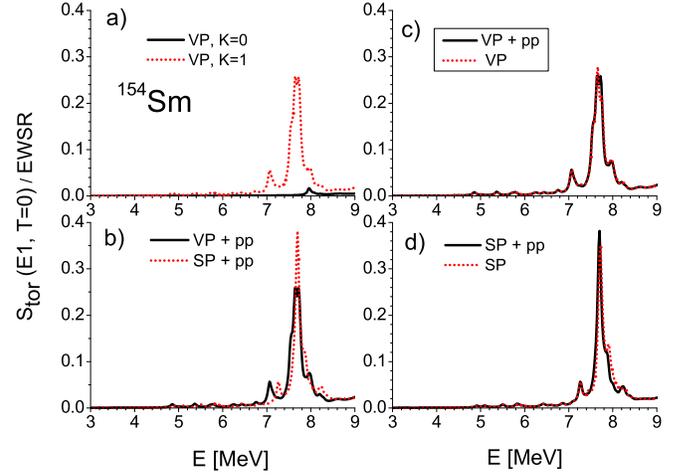}}
\vspace{0.3cm}
\caption{The calculated low-energy isoscalar toroidal E1 strength in
  $^{154}$Sm.  (a) VP with pp channel: $K=0$ and $K=1$ components, b)
  VP and SP with pp-channel, c) VP with and without pp-channel, d) SP
  with and without pp-channel.}
\label{fig:10}
\end{figure}

Fig. \ref{fig:7} shows the QRPA isoscalar dipole strength
(ISGDR) as a fraction of the computed energy-weighted sum rule
EWSR(E1,T=0)=$\int dE S_{\rm com}(E1, T=0; E)$ (a similar presentation
but in terms of experimental EWSR is used for experimental data
\cite{CDRexp}). We see that, in agreement with experiment, the
computed ISGDR includes two broad bumps at 10-17 and 17-40 MeV. Similar
bumps were also observed in other (spherical) nuclei \cite{CDRexp,Young90Zr,Uch04}.
They are often treated as TDR and CDR respectively,
see e.g. \cite{CDRexp}. However, our previous studies
\cite{Re13,Kv13Sm,Kv14Yb,Ne17,NeDre14} for different nuclei and
present analysis for $^{154}$Sm suggest that both broad bumps in ISGDR
belong to CDR. Comparison of Figs. \ref{fig:7} and \ref{fig:9}
shows that even spikes at 4--6 MeV can hardly be considered as
being toroidal because they are almost absent in the toroidal
distributions shown in Figs. \ref{fig:9} and \ref{fig:10}.

Fig. \ref{fig:7} demonstrates that strengths obtained with VP and SP are almost identical.
The effect of the pp-channel for the strength at $E >$ 10 MeV is
negligible (not shown).
Concerning the deformation effects, $K=0$ and $K=1$ branches show the
familiar sequence (like in GDR) at E=10-40 MeV,   namely that the $K=0$ branch
lies below the $K=1$ branch.

The ISGDR strength at lower energies is shown in Fig. \ref{fig:8}. Plot a) demonstrates
that $K=1$ strength dominates at E=8--10 MeV but becomes weaker at 3--8 MeV.
So, at 3--8 MeV, the ratio of $K=0$ and $K=1$
strengths is opposite to that in PRD.  Plots b)-d) of
Fig. \ref{fig:8}) also show very close results for VP as compared to SP
and an almost negligible effect of the pp-channel.

\begin{figure}
\resizebox{0.48\textwidth}{!}{%
\includegraphics{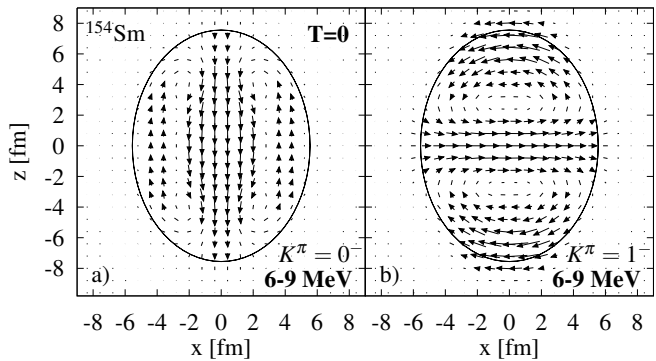}}
\vspace{0.3cm}
\caption{The isoscalar QRPA current transition densities in x-z plane
  for dipole excitations at 6--9 MeV in $^{154}$Sm: a) $K=0$, b)
  $K=1$. The currents are obtained for VP. Magnitude of the
  current is determined by arrow length (in arbitrary units). The
  nuclear surface is indicated by the solid line.}
\label{fig:11}
\end{figure}

Fig. \ref{fig:9} shows the toroidal isoscalar strength function
obtained with the transition operator (\ref{15}) as a fraction of the
integral toroidal strength (EWSR). We see that the strength has a
broad plateau at 10--40 MeV. Most probably such a broad distribution
is caused by the coupling between TDR and CDR \cite{Kv11}. In this
plateau, the $K=1$ strength dominates at $E >$ 25 MeV.  Below, at
12--25 MeV, the $K=0$ and $K=1$ strengths are about the same. Even
lower, at $E <$ 12 MeV, the $K=1$ strength becomes
strongly dominant. In particular a strong toroidal peak at 7--8 MeV is
almost fully built from $K=1$ strength. Note that SP makes this peak
essentially higher than VP.  Perhaps this happens because the toroidal $K$=1
motion is strongest just at the nuclear surface, see Fig. \ref{fig:11}
below. So the description of the TDR peak
depends on the sort of the pairing. This is the only place where
we could observes a significant difference between VP and SP.

The TDR strength is presented in more detail in
Fig. \ref{fig:10}. Plot b) demonstrates a significant difference
between VP and SP results for the highest TDR peak. Plots c) and d)
show a negligible contribution of the pp-channel. Further, plot a)
confirms that the TDR strength is almost fully provided by $K$=1
contribution. This peculiarity of the TDR corresponds to the PDR
picture (where $K$=1 strength also dominates at $E <$ 8 MeV) but is
opposite to low-energy part of the CDR with dominant $K$=0
strength. It is interesting that the PDR peaks roughly correspond to
the PDR structures at 7--8 MeV but have no counterpart in the
low-energy CDR located at 4-6 MeV. This is one more indication
for the intimate
relation between TDR and PDR, discussed in \cite{Re13}.  Note that
the peaked TDR strength appears even in the unperturbed (without
residual interaction) strength function \cite{NeDre14}. The isoscalar
residual interaction only downshifts the structure by $\sim$ 2 MeV.

To illustrate the mainly toroidal nature of dipole states at 6--9 MeV,
we show in Fig. \ref{fig:11} their isoscalar average current
transition densities (CTD) for the branches $K=0$ and $K=1$. The
method of calculation of the average CTD in QRPA is described in
detail in \cite{Re13}. Remind that CTD depend only on the
structure of QRPA states and are not affected by the probing
external field. In Fig. \ref{fig:11}, we see for both $K=0$ and $K=1$
basically toroidal flow. The toroidal axes are directed along z-axis
for $K=0$ and x-axis for $K=1$. It is interesting that the toroidal
flow resembles a vortex-antivortex configuration. Fig.
\ref{fig:11} shows that for $K=1$ the flow is much stronger,
especially at the nuclear surface. This supports the previous finding
that the TDR is mainly realized in the $K=1$ branch. Perhaps, the
nuclear geometry in the $K=1$ case is more suitable for hosting the
toroidal flow and thus makes the flow more energetically favorable.  Our
analysis shows that the strong $K=1$ flow at the nuclear surface is
mainly provided by neutrons. The strong motion of the neutron
  surface against
the remaining core corresponds to a typical PDR picture. So
Fig. \ref{fig:11} confirms our previous suggestion \cite{Re13}
that PDR is mainly a local boundary manifestation of the toroidal
nuclear flow.
\begin{figure}
\resizebox{0.48\textwidth}{!}{%
\includegraphics{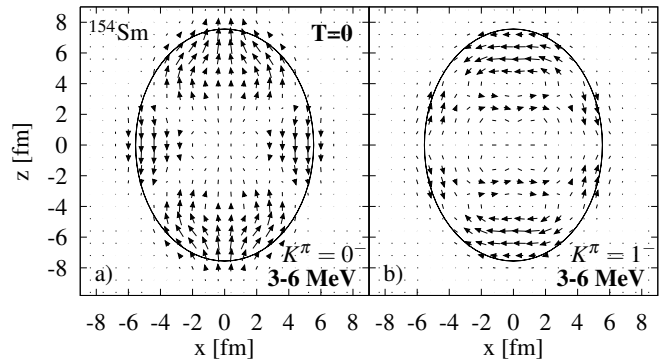}}
\vspace{0.3cm}
\caption{The same as in Fig. 11 but
  for dipole excitations at 3--6 MeV.}
\label{fig:12}
\end{figure}

Finally, Fig. \ref{fig:12} shows the currents for the dipole excitations
at 3--6 MeV (exhibited at Fig. 8a). The flow considerably
deviates from the toroidal picture.

Altogether, Figs. \ref{fig:5}--\ref{fig:12} show that in the dipole
resonances (GDR, PDR, TDR, CDR): i) the pp-channel is negligible; ii)
the difference between VP and SP results is small (except for the
strongest TDR peak at 7--8 MeV); iii) at $E>$ 10 MeV, the familiar
sequence of $K=$0 and $K=$1 branches is observed; iv) at $E<$ 10 MeV,
the ratio between the branches depends on the resonance ($K=1$
strength dominates in TDR and PDR but becomes minor in CDR); v) The
$K=1$ strength forms an impressively narrow low-energy peak in the
TDR (the latter feature may serve as a fingerprint of the TDR in
future experiments);  vi) current transition densities (CTD) confirm
the mainly toroidal nature of dipole excitations at 6--9 MeV.

Figs. \ref{fig:5}--\ref{fig:12} allow to do
some additional comments concerning origin of PDR and ISGDR, deformation
splitting of these resonances, and perspectives of experimental
observation of TDR.

1) The TDR operator (\ref{15}) has the radial dependence $\sim r^2$
and so surface reactions should be sensitive probes of this resonance.
In particular, the ($\alpha,\alpha'\gamma$) reaction in deformed nuclei
looks suitable to excite the isoscalar TDR and determine, using Alaga
rules \cite{Ala55}, a dominant K=1 low-energy TDR peak.
The ratio of intensities of $E1K$-transitions from a given dipole QRPA
$K$-state to the members $I^{\pi}K=0^+0$ and $2^+0$ of the ground-state
rotational band is determined through Clebsh-Gordan coefficients as
\begin{equation}
 A_K=|\frac{\langle 1K 1 -K |00\rangle}{\langle 1K 1-K|20\rangle}|^2.
\end{equation}
This yields $A_0$=0.5 and $A_1$=2. Thus $K=0$ and $K=1$ branches
can be distinguished. Although these dipole transitions are isoscalar, they should
be strong enough to be detectable since the peaked TDR is a collective mode
\cite{Kv14Yb,Ne17,NeDre14}. Note that in $^{154}$Sm the TDR peak lies below
the neutron and proton threshold energies, which should favor its $\gamma$-decay.

2) The relevance of the ($\alpha,\alpha'\gamma$) reaction to observe the TDR
is not so evident. Note that, despite the TDR/CDR relation,
the irrotational ISGDR strength in Figs. \ref{fig:7}-\ref{fig:8}
does not match the vortical toroidal strength exhibited in Figs.
\ref{fig:9}-\ref{fig:10}.
There is a general question if the ($\alpha,\alpha'$) reaction
is, in principle, able to probe vortical modes. We need here further analysis.
From the theoretical side, DWBA calculations for ($\alpha,\alpha'$) angular
distributions should be performed, using QRPA input formfactors for the
TDR low-energy $K=1$ peak.

3) Following our results, the ISGDR observed at 10-33 MeV in $(\alpha,\alpha')$ scattering
to small angles \cite{CDRexp} is perhaps a fully irrotational mode dominated
by compression-like dipole excitations. In other words, the observed ISGDR
does not include the TDR at all. Indeed both observed broad bumps at 10-17 MeV and
17-33 MeV are located higher than the TDR peak predicted at 7-8 MeV. Moreover,
in this experiment, only excitations at $E>$ 10MeV are measured, i.e. the actual TDR
energy region is fully omitted. In the same experiment for other nuclei
($^{116}$Sn, $^{144}$Sm and $^{208}$Pb) the TDR low-energy region was also missed
\cite{CDRexp}. At the same time, some low-energy peaks (which can be attributed to the TDR)
were measured in $^{90}$Zr, $^{116}$Sn, $^{144}$Sm and $^{208}$Pb in
an alternative
$(\alpha,\alpha')$ experiment \cite{Uch04} where the interval $E>$ 8 MeV was inspected.
These peaks can match a part of the TDR. However we still need  measurements  for
the lower energy interval which fully embraces the predicted TDR.

4) As discussed in \cite{Re13}, the PDR is probably a complicated mixture of various dipole
excitations (toroidal, compression, GDR tail) with a dominant contribution of the toroidal
fraction. Following Figs. \ref{fig:6}, \ref{fig:8}, \ref{fig:10}, the dipole response
crucially depends on the probe. The PDR structure  in deformed nuclei (K-branches) also
should depend on the reaction applied. For example, following Figs. \ref{fig:8} and \ref{fig:10},
$K$-branches of the PDR can arise separately as responses to different probes: $K$=0 in
the CDR/ISGDR response and $K=1$ in the toroidal response. If so, then investigation of
CDR/ISGDR and TDR could be useful to understand the deformation splitting of the PDR.

\section{Conclusions}
\label{concl}

Effects of nuclear deformation and monopole pairing in nuclear
excitations with $K^{\pi} \ne 0^+$ were investigated within the
self-consistent Quasiparticle Random Phase Approximation (QRPA) method
\cite{Re16} using the Skyrme parametrization SLy6
\cite{Cha97}. Mean field and pairing were treated within the HF+BCS
scheme. As representative examples the axially deformed  nuclei
$^{152,154,156}$Sm were considered.  One of the main aims was to investigate
the role of the particle-particle (pp) channel in the residual QRPA
interaction. This channel is known to be important for $K^{\pi} = 0^+$
states \cite{SSS,Ri80} but its role in $K^{\pi} \ne 0^+$ excitations
was yet unclear. We scrutinized this problem for the lowest
$\gamma$-vibrational $K^{\pi}=2^+$ states and for dipole low- and
high-energy excitations: isovector giant dipole resonance (GDR),
isovector pygmy dipole resonance (PDR), toroidal dipole resonance
(TDR), and compression dipole resonance (CDR). We also compared QRPA
results for these excitations with volume and surface
pairing. Last but not least, we analyzed deformation effects in
the dipole resonances, especially in the low-energy interval where PDR
is located. The following results have been obtained:
\\
1) The effect of the pp-channel in $K^{\pi} \ne 0^+$
excitations is negligible. This can be explained by the absence of the
proton-neutron part in the pairing forces (though the
proton-neutron interaction dominates in the ph-channel).
As a result, the pp-channel turns
out to be much weaker than its ph-counterpart and finally can be safely
neglected in description of $K^{\pi} \ne 0^+$ modes. This result is important
from the practical side because omitting the pp-channel considerably
simplifies QRPA calculations, which is especially beneficial for
deformed nuclei where QRPA calculations are time-consuming.  Since the
effect of the pp-channel is fairly small, its omission does not spoil the
consistency of QRPA.  In our study, this result was obtained for
low-energy $K^{\pi}=2^+$ states and dipole resonances. However it is
  likely to be valid for any nuclear excitation with
$K^{\pi} \ne 0^+$ (dipole, quadrupole, octupole).
\\
2) Our calculations have shown that, as a rule,  the difference between
  the QRPA results obtained with from volume pairing (VP) and from surface pairing
  (SP) is small. A noticeable difference was found
  only for the collective low-energy peak of the TDR.
\\
3)  We analyzed deformation effects arising due to competition
between $K=0$ and $K=1$ branches in the dipole excitations,
comparing the behavior of the branches in various dipole
resonances (GDR, PDR, TDR, CDR).  It was shown that, in all the
resonances, the familiar sequence of $K=0$ and $K=1$ branches takes
place for excitation energy $E >$ 10 MeV, namely the $K=0$
structure lies lower than the $K=1$ one as it should be in
prolate nuclei.  However, at lower excitation energies $E<$ 10 MeV,
the behavior of the branches depends on the resonance type. In
particular, in the PDR and TDR, the $K=1$ strength becomes
dominant. For TDR, this effect manifests itself in the appearance of
very strong peak at $\sim$ 7--8 MeV, which is almost completely built
from $K=1$ strength. This remarkable feature can be used as a
fingerprint of the TDR in future experiments. The ($\alpha,\alpha'\gamma$)
reaction in deformed nuclei looks most promising. The peaked $K=1$ strength
can be identified inspecting intensities of the  $\gamma$-decay
in terms of Alaga rules.
%
\\
4)  Following our results, the ISGDR observed in $(\alpha,\alpha')$ scattering
to forward angles \cite{CDRexp} is basically irrotational and perhaps does not include
at all the low-energy vortical TDR. To search the TDR, the lower excitation energies
should be experimentally inspected.
\\
5)The nuclear flow in the PDR energy region is basically toroidal.
The deformation structure of PDR can be related to $K$ branches in TDR and CDR.

\vspace*{0.2cm}

The work was partly supported by Votruba-Blokhincev (Czech Republic-BLTP JINR)
and Heisenberg-Landau (Germany - BLTP JINR) grants.
The support of the Czech Scinece Foundation Grant Agency (project P203-13-07117S)
is appreciated. A.R. is grateful for support from Slovak Research and Development
Agency under Contract No. APVV-15-0225.

\appendix  
\numberwithin{equation}{section}
\section{Skyrme functional}
\label{app-A}

The Skyrme energy functional (\ref{Efunc}) has the kinetic-energy,
Skyrme, Coulomb and pairing parts \cite{Be03,Rei92d} with ($q=n,p$)
\begin{equation}
\mathcal{E}_{\rm kin} = \int\mathrm{d}\vec{r}
\sum_q \frac{\hbar^2}{2m_q}\tau_q(\vec{r}) ,
\label{A.2}
\end{equation}
\begin{eqnarray}
&&
\mathcal{E}_{\rm Sk}  = \int\mathrm{d}\vec{r}
\bigg\{ \frac{b_0}{2}\rho^2 - \frac{b_0'}{2}\sum_q\rho_q^2
+\frac{b_2}{2}(\vec{\nabla}\rho)^2
\label{A.3}
\\
&&
- \frac{b_2'}{2}\sum_q(\vec{\nabla}\rho_q)^2 + b_1(\rho\tau\!-\!\vec{j}^{\,2})
- b_1'\sum_q(\rho_q\tau_q\!-\!\vec{j}_q^{\,2})
\nonumber \\
&&
 +\tilde{b}_1\Big(\vec{s}\cdot\vec{T} -  {\bf J}^2\Big)
 +\tilde{b}_1'\sum_q\Big(\vec{s}_q\cdot\vec{T}_q-{\bf J}_q^2 \Big)
\nonumber \\
&&
+\frac{b_3}{3}\rho^{\alpha+2}-\frac{b_3'}{3}\rho^\alpha\sum_q\rho_q^2
-b_4\big[\rho\vec{\nabla}\cdot\vec{J}
+\vec{s}\cdot(\vec{\nabla}\times\vec{j})\big]
\nonumber
\end{eqnarray}
\begin{eqnarray}
&&
-b_4'\sum_q\big[\rho_q\vec{\nabla}\cdot\vec{J}_q+\vec{s}_q\cdot (\vec{\nabla}\times\vec{j}_q)\big]
\nonumber\\
&&
+\frac{\tilde{b}_0}{2}\vec{s}^2-\frac{\tilde{b}_0'}{2}\sum_q\vec{s}_q^{\,2}
+\frac{\tilde{b}_2}{2}\sum_{ij}(\nabla_i s_j)^2
\nonumber \\
&&
-\frac{\tilde{b}_2'}{2}\sum_q \sum_{ij}(\nabla_i s_j)_q^2
+\frac{\tilde{b}_3}{3}\rho^\alpha\vec{s}^{\,2}
-\frac{\tilde{b}_3'}{3}\rho^\alpha\sum_q \vec{s}_q^{\,2} \bigg\}\: ,
\nonumber
\end{eqnarray}
\begin{eqnarray}
\label{A.4}
&&
\mathcal{E}_{\rm Coul} =
\frac{1}{2} e^2 \int \mathrm{d}\vec{r}
\int \mathrm{d}\vec{r}' \frac{\rho_p(\vec{r}) \rho_p(\vec{r}')}
{|\vec{r} - \vec{r}'|}
\\
&& \qquad \qquad \qquad
- \frac{3}{4}\Big(\frac{3}{\pi}  \Big)^{\frac{1}{3}} e^2
\int \mathrm{d}\vec{r} \rho_p(\vec{r})^{\frac{4}{3}} .
\nonumber
\end{eqnarray}
The pairing part is given in Eq. (\ref{Epair}) of Section 2.2.
The functional depends on time-even ($\rho$, $\tau$, ${\bf J}$,
$\tilde\rho$) and time-odd ($\vec{j}$, $\vec{s}$,
$\vec{T}$) local densities and currents. The total densities
are sums of the proton and neutron parts, e.g. $\rho=\rho_n + \rho_p$.
In HF+BCS approach, the densities of our main interest, nucleon $\rho$ and pairing
$\tilde\rho$, have the form (\ref{rho}) and (\ref{rho_pair}). The relation of the
Skyrme parameters $b_i$ with the standard ones $t_i,\:x_i$ can
be found in \cite{Re16,Be03,Rei92d}

\section{HF+BCS equations}
\label{app-b}

The zero-range monopole pairing is defined by Eqs. (\ref{Epair})-(\ref{Gpair}).
It results in the BCS equations:
\begin{equation}
[u_i^q]^2=
  \frac{1}{2} \: \left\{1 + \frac{e_i^q -
  \lambda_q}{\sqrt{\left[e_i^q - \lambda_q\right]^2
  + [\Delta_i^{q}]^2}} \right\},
\label{u_i}
\end{equation}
\begin{equation}
 [v^q_i]^2 =
  \frac{1}{2} \: \left\{1 - \frac{e_i^q -
  \lambda_{q}}{\sqrt{\left[e_i^q - \lambda_{q}\right]^2
  + [\Delta_i^{q}]^2}} \right\},
\label{v_i}
\end{equation}
\begin{equation}
  \Delta_i^{q}
  = -  \sum_{j \in q}^{K>0}
  f_{j}^q V^{(\rm{pair}, q)}_{i \bar{i} j \bar{j}} \: v_{j}^q  u_{j}^q ,
\label{Del_i}
\end{equation}
\begin{equation}
  N_{q}
  =
  \sum_{i \in q}^{K>0} f_{i}^q
  \left\{1 - \frac{e_i^q - \lambda_{q}}{\sqrt{\left[e_i^q -
         \lambda_{q}\right]^2 + [\Delta_i^{q}]^2}} \right\} ,
\label{Npair}
\end{equation}
where
$e_i^q$ is the s-p energy, $u_{i}^q, v_{i}^q$ are
Bogoliubov coefficients, $\Delta_i^{q}$ is the pairing gap, $\lambda_{q}$ is
the chemical potential, $f_{i}^q$ is the cut-off function (\ref{f-cut}),
$N_p=Z$ and $N_n=N$ are proton and neutron numbers, and
$V_{i \bar{i} j \bar{j}}^{(\rm{pair, q})}
=\langle i\bar{i}|V^q_{\rm pair}(\vec{r},\vec{r}')|j\bar{j}\rangle$
is the pairing matrix element.

The total HF+BCS Hamiltonians consists from the local pairing and HF parts:
\begin{equation}
{\hat h}^q_{\rm HF+BCS}(\rho,{\tilde \rho})=
{\hat h}^q_{\rm pair}(\rho,{\tilde \rho})
+  {\hat h}^q_{\rm HF} (\rho,\tau,{\bf J},\tilde{\rho}) .
\end{equation}
The pairing part
\begin{equation}
{\hat h}^q_{\rm pair}(\rho,{\tilde \rho})(\vec{r})=
\frac{\delta \mathcal{E}_{\rm pair}}
{\delta {\tilde \rho_q}(\vec{r})}
=\frac{1}{2}{\tilde \rho_q}(\vec{r})G_q(\vec{r})
\label{hpair}
\end{equation}
depends on the pairing density $\tilde\rho$ and, in SP case,
on total nucleon density $\rho$ (through the factor $G_q(\vec{r})$).

The mean-field part
\begin{equation}
{\hat h}^q_{\rm HF} (\rho, \tau,{\bf J},\tilde\rho)(\vec{r})=
\frac{\delta \mathcal{E}}{\delta \rho_q(\vec{r})}
+\frac{\delta \mathcal{E}}{\delta \tau_q(\vec{r})}
+\frac{\delta \mathcal{E}}{\delta \mathbf{J}_q(\vec{r})}
\label{hHF}
\end{equation}
is determined by the time-even densities involved to the Skyrme functional
(\ref{A.3}). For SP, the Hamiltonian (\ref{hHF}) gains dependence on the pairing density
$\tilde\rho$.

BCS equations (\ref{u_i})-(\ref{Npair}) are coupled to the density-dependent
HF equations
\begin{equation}
\label{HF}
  {\hat h}^q_{\rm HF} (\rho,\tau,{\bf J},\tilde{\rho})(\vec{r})
   \psi^q_i(\vec{r}) =e_i^q \psi^q_i(\vec{r})
\end{equation}
via the pairing $u_{i}^q, v_{i}^q$-factors in the densities, see e.g. Eq. (\ref{rho})
for $\rho_q (\vec{r})$. In turn,
the BCS equations are affected by the HF mean field
through single-particle energies $e_i^q$ and wave functions $\psi_i^q$.
In our HF+BCS scheme, BCS equations (\ref{u_i})--(\ref{Npair})
are solved at each HF iteration \cite{Skyax}.

\section{QRPA with pp-channel}
\label{app-D}

The QRPA equations read \cite{Ri80}
\begin{equation}
\left(
\begin{array}{cc}
A  &  B  \\
B^* & A^*
\end{array}
\right)
\left(
\begin{array}{c}
X^{(\nu)} \\
Y^{(\nu)}
\end{array}
\right)
=
\hbar E_{\nu} \:
\left(
\begin{array}{cc}
1  &  0 \\
0  & -1
\end{array}
\right)
\left(
\begin{array}{c}
X^{(\nu)} \\
Y^{(\nu)}
\end{array}
\right)
\label{1}
\end{equation}
where $X^{(\nu)}_{ij}$ and $Y^{(\nu)}_{ij}$ are forward and backward amplitudes
of QRPA phonon creation operators $Q^+_{\nu}$
\begin{equation}
Q^+_{\nu} = \frac{1}{2}
\sum_{ij}
\big( X^{(\nu)}_{ij} \: \alpha^+_i \alpha^+_j - Y^{(\nu)}_{ij} \: \alpha_j \: \alpha_i \big) ,
\label{2}
\end{equation}
$E_{\nu}$ is the phonon energy, $\alpha^+_i (\alpha_j)$ is the quasiparticle creation
(annihilation) operator. The RPA matrices $A$ and $B$ have the form
\begin{eqnarray}
A_{\omega \omega'} &=& \epsilon_{\omega} \delta_{\omega, \: \omega^{'}}
\nonumber
\\
&+& \sum_{d\:d^{\:'}} \int \!\! \mathrm{d}\vec{r} \! \int \!\! \mathrm{d}\vec{r}' \frac{\delta^2 \mathcal{E}}
{\delta \mathcal{J}^d(\vec{r}) \: \delta \mathcal{J}^{d'}(\vec{r^{'}})} \:
\delta \mathcal{J}^d_{\omega}(\vec{r}) \: \delta \mathcal{J}^{d'}_{\omega'}(\vec{r}')
\nonumber \\
B_{\omega \omega^{'}} &=& - \:
 \sum_{d\:d^{\:'}} \int \!\! \mathrm{d}\vec{r}\! \int \!\! \mathrm{d}\vec{r}' \frac{\delta^2 \mathcal{E}}
{\delta \mathcal{J}^d(\vec{r}) \: \delta \mathcal{J}^{d'}(\vec{r^{'}})} \:
\delta \mathcal{J}^d_{\omega}(\vec{r}) \: \delta \mathcal{J}^{d'}_{\omega'}(\vec{r}')
\nonumber
\label{9}
\end{eqnarray}
where  $\omega = ij,\: \bar{i} \bar{j},\: i \bar{j} \in q$ and
$\omega' = i'j',\: \bar{i}' \bar{j}',\: i' \bar{j}' \in q'$ ($i > j$);
$\delta_{ij,\: i^{'} j^{'}} \equiv \delta_{i i^{'}} \delta_{j j^{'}} -
\delta_{i j^{'}} \delta_{j \:i^{'}}$, $\delta_{i\: \bar{j},\: i^{'} \bar{j^{'}}}
\equiv \delta_{i i^{'}} \delta_{j j^{'}}$, $\delta_{ij,\: i^{'} \bar{j^{'}}} = 0$.
Further, $\epsilon_{\omega} \equiv \epsilon_i + \epsilon_j$ is 2qp
energy;
$\delta \mathcal{J}^d_{\omega}(\vec{r})=\langle \omega |\hat{\mathcal{J}}^d|0\rangle$
is the transition density for the transition between 2qp state $|\omega\rangle$
and BCS vacuum $|0\rangle$, $\hat{\mathcal{J}}^d$ is the density operator in 2qp representation
(in terms of $\alpha^+_i\alpha^+_j, \: \alpha_i\alpha_j$);  the set $\mathcal{J}^d$ includes
the densities and currents $d \in \:\rho,\: \tau,\: {\bf J}, \: \vec{j},
\: \vec{s} , \: \vec{T}, \tilde{\rho}$. For more details see \cite{Re16}.

The pairing functional $\mathcal{E}_{\rm pair}$ (\ref{Epair}) leads to additional terms
in the residual interaction, which constitute the pp-channel:
\begin{equation}
\frac{\delta^2 \mathcal{E}} {\delta \rho(\vec{r}) \: \delta \rho (\vec{r}^{'})},
\:
\frac{\delta^2 \mathcal{E}} {\delta \rho(\vec{r}) \: \delta \tilde{\rho} (\vec{r}^{'})},
\:
\frac{\delta^2 \mathcal{E}} {\delta \tilde{\rho} (\vec{r}) \: \delta \rho (\vec{r}^{'})},
\:
\frac{\delta^2 \mathcal{E}} {\delta \tilde{\rho}(\vec{r}) \: \delta \tilde{\rho} (\vec{r}^{'})} .
\nonumber
\end{equation}
In SP, all the terms take place while in VP only the last term remains.
In the QRPA matrices $A$ and $B$, the pp-channel is presented by the terms:
\begin{eqnarray}
&&A^{\rm pair}_{qq'}(i j, i' j') =
-\frac{\eta \: \gamma \: (\gamma -1)}{4\: \rho^{\gamma}_0}
\label{C.1}
\\[-0.2cm]
&\cdot& \sum_{q=n,p} V_q \:
\int \mathrm{d}\vec{r} \: \tilde{\rho}^2_q(\vec{r}) \:\: \rho^{\gamma -2}(\vec{r})
\: \delta \rho_{ij}(\vec{r}) \: \delta \rho_{i'j'}(\vec{r}) ,
\nonumber \\
&&B^{\rm pair}_{qq'}(i j, i' j') = -\: A^{\rm pair}_{qq'}(i j, i' j') ,
\label{C.2}
\end{eqnarray}
\begin{eqnarray}
&& \:
A^{\rm pair}_{qq'}(i \bar{j}, i' \bar{j}')
= -\frac{\eta \: \gamma \: (\gamma -1)}{4\: \rho^{\gamma}_0}
\label{C.3} \\
&\cdot& \sum_{q=n,p} V_q \: \int \mathrm{d}\vec{r} \: \tilde{\rho}^2_t(\vec{r}) \:\: \rho^{\gamma -2}(\vec{r})
\delta \rho_{ij}(\vec{r}) \: \delta \rho_{i'j'}(\vec{r})
\nonumber \\
&&  +
\frac{1}{2} \: \delta_{q q'} V_q \: \int \mathrm{d}\vec{r}
\Big[1 - \eta \: \Big(\frac{\rho(\vec{r})}
{\rho_0}\Big)^{\gamma} \Big]
\delta \tilde{\rho}_{ij}(\vec{r}) \: \delta \tilde{\rho}_{i'j'}(\vec{r})
\nonumber \\
&& -
\frac{\eta \: \gamma}{2 \rho_0^{\gamma}} \: V_q \: \int \mathrm{d}\vec{r} \:\tilde{\rho}_q(\vec{r})
\: \rho^{\gamma-1}(\vec{r})
\delta \tilde{\rho}_{ij}(\vec{r}) \: \delta \rho_{i'j'}(\vec{r})
\nonumber \\
&& -
\frac{\eta \: \gamma}{2 \rho_0^{\gamma}} \: V_{q^{'}} \: \int \mathrm{d}\vec{r} \:
\tilde{\rho}_{q^{'}}(\vec{r}) \: \rho^{\gamma-1}(\vec{r})
\delta \rho_{ij}(\vec{r}) \: \delta \tilde{\rho}_{i'j'}(\vec{r})
\nonumber\\
&&B^{\rm pair}_{qq'}(i \bar{j}, i' \bar{j}') = -\: A^{\rm pair}_{qq'}(i \bar{j}, i' \bar{j}') ,
\label{C.4}
\end{eqnarray}
The contributions (\ref{C.1})-(\ref{C.2}) and the first term in (\ref{C.3}) are
zero for both VP ($\eta$=0) and SP (if $\gamma$=1).
The last two cross terms in (\ref{C.3}) exist only for the density-dependent SP (when
$\eta \ne$ 0).
Only the second term in (\ref{C.3}), including merely pairing transitions densities
$\delta \tilde{\rho}$, exists for both VP and SP.
 Expressions (\ref{C.1})-(\ref{C.4}) correspond to the formalism derived
in \cite{Te05} for spherical nuclei.

Expressions (\ref{C.1})-(\ref{C.4}) include contributions to pp-channel from both diagonal ($i=j$)
and non-diagonal ($i \ne j$) 2qp configurations. Note that, in the quasiparticle-phonon model (QPM)
with the constant pairing \cite{SSS}, only diagonal contributions are taken into account, thus
the pp-channel exists only for $K^{\pi}=0^+$ states. In our study for states with
$K^{\pi} \ne 0^+$, the selection rules allow only non-diagonal configurations.

\end{document}